\RequirePackage{fix-cm}
\documentclass[smallextended]{svjour3}       
\smartqed  
\usepackage{graphicx}
\def \be {\begin{equation}}
\def \en {\end{equation}}

\begin{document}

\title{The AGILE Data Center and its Legacy
}


\author{Carlotta Pittori, \\
on behalf of the AGILE-SSDC Team}


\institute{C. Pittori \at
ASI-SSDC, via del Politecnico, I-00133 Roma, Italy \\
INAF-OAR, Via Frascati 33, I-00078 Monte Porzio Catone (RM), Italy\\
              \email{carlotta.pittori@ssdc.asi.it}  \\
ORCID iD: 0000-0001-6661-9779
}

\date{Received: date / Accepted: date}

\maketitle

\begin{abstract}
We present an overview of the main AGILE Data Center activities
and architecture.
AGILE is a space mission of the Italian Space Agency (ASI)
in joint collaboration with INAF, INFN, CIFS, and with the 
participation of several Italian 
space industry companies.
The AGILE satellite was launched on April 23, 2007, and
is devoted to the observation of the gamma-ray Universe
in the 30 MeV -- 50 GeV energy range, with simultaneous X-ray imaging 
capability in the 18--60 keV band.
The AGILE Data Center, part of the ASI multi-mission 
Space Science Data Center (SSDC, previously known as ASDC) 
is in charge of all the scientific operations: data management, archiving, 
distribution of AGI\-LE data and scientific software, and user support.
Thanks to its sky monitoring capability and fast ground segment alert system,
AGILE is substantially improving our knowledge of the gamma-ray sky,
and provides a crucial contribution
to multimessenger follow-up of gravitational waves and neutrinos.
\end{abstract}

\section{Introduction}
\label{intro}

AGILE (Astrorivelatore Gamma ad Immagini
LEggero) is an Italian scientific space mission for high
energy astrophysics funded by the Italian Space Agency (ASI)
with scientific and programmatic participation by INAF, INFN, CIFS,
several Italian universities and industrial contractors.
The AGILE very innovative instrument combines for the first time
a gamma-ray imager based on solid-state silicon technology, 
and a hard X-ray imager.
The AGILE satellite was successfully launched
on 2007 April 23rd from the Indian
base of Sriharikota, and it was inserted in an
equatorial low Earth orbit with an inclination of 2.5 degrees 
and average altitude of 535 km. 

\begin{figure*}
\vskip -4. truecm
\centerline{\resizebox{100mm}{!}{\includegraphics{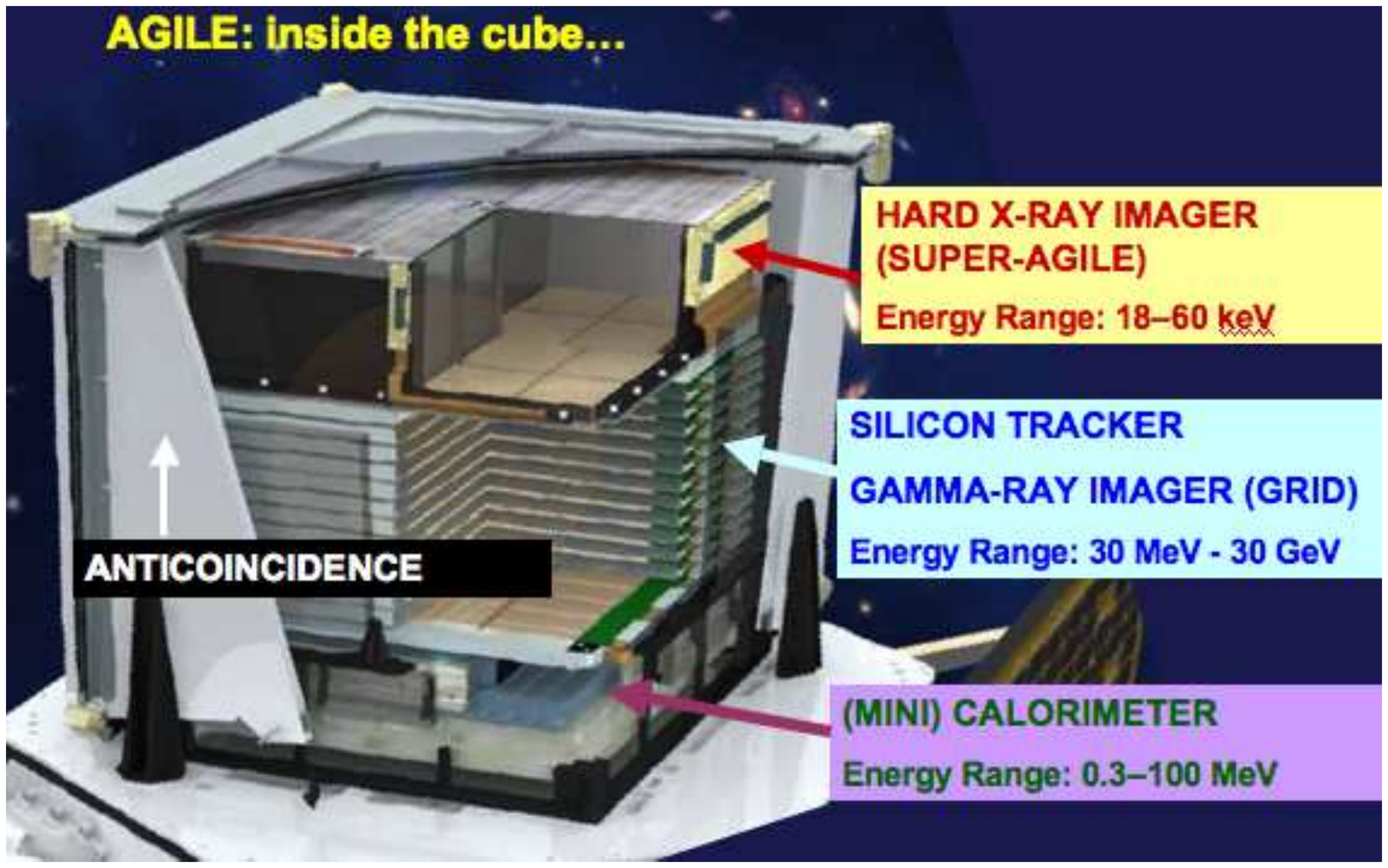}}}
\vskip -4. truecm
\caption{The AGILE payload.}
\label{fig01}
\end{figure*}

The AGILE instrument shown in Fig. \ref{fig01}
is a cube of 60 cm side, weighting only about 100 kg. It consists of two detectors 
using silicon technology: the gamma-ray imager GRID 
and the hard X-ray detector SuperAGILE 
for the simultaneous detection and imaging of photons in the 30 MeV - 50 GeV 
and in the 18 - 60 keV energy ranges. 
The payload is completed by two non-imaging detectors: a Mini-Calorimeter (MCAL)
sensitive in the energy range 350 keV to 100 MeV, 
and an anti-coincidence system. 
Further details on the AGILE scientific mission are reported in  \cite{Tavani2009}
and Tavani (2019), these Proceedings.

The AGILE satellite, designed to achieve a
nominal lifetime of two years, has completed more than
ten years of operations in orbit, and it is substantially contributing 
to improve our knowledge of the high-energy sky, 
also providing an important contribution
to multimessenger follow-up of gravitational waves and neutrinos.
Satellite operations and all payload functions are nominal and
AGI\-LE continues its mission with high efficiency.

In this paper we describe the architecture and functionalities of the
AGI\-LE Data Center, which is part of the ASI multi-mission 
Space Science Data Center (SSDC, previously known as ASDC).
In Sect. \ref{ADC} we present the general architecture of the
AGILE Data Center, and in Sect. \ref{FLOW} the AGILE data flow.
We describe in Sect. \ref{obsmodes} the AGILE observation modes, in
Sect. \ref{PROCESSING} the data processing system and data 
levels, in Sect. \ref{agileservices} the ``AGILE Services'' and the AGILE storage,
in Sect. \ref{datapolicy} the data policy and distribution, and 
in Sect. \ref{catalogs} the AGILE published catalogs and 
their interactive webpages. Finally in Sect. \ref{LV3}
we present the AGILE legacy archive and the AGILE-LV3 web tool for 
scientific analysis.


\section{The AGILE Data Center}
\label{ADC}

The AGILE Data Center (ADC)\footnote{https://agile.ssdc.asi.it/}
is in charge of all the scientific oriented activities 
related to the analysis, archiving and distribution of AGILE data.
It is part of the ASI Space Science Data Center (SSDC) 
located at the ASI Headquarters in Rome (Italy),
and it includes scientific personnel from both the SSDC and the AGI\-LE Team, 
with the support of the IT technical Team of the  multi-mission ASI data center.

\begin{figure*}
\vskip -4.7 truecm
\centerline{\resizebox{110mm}{!}{\includegraphics{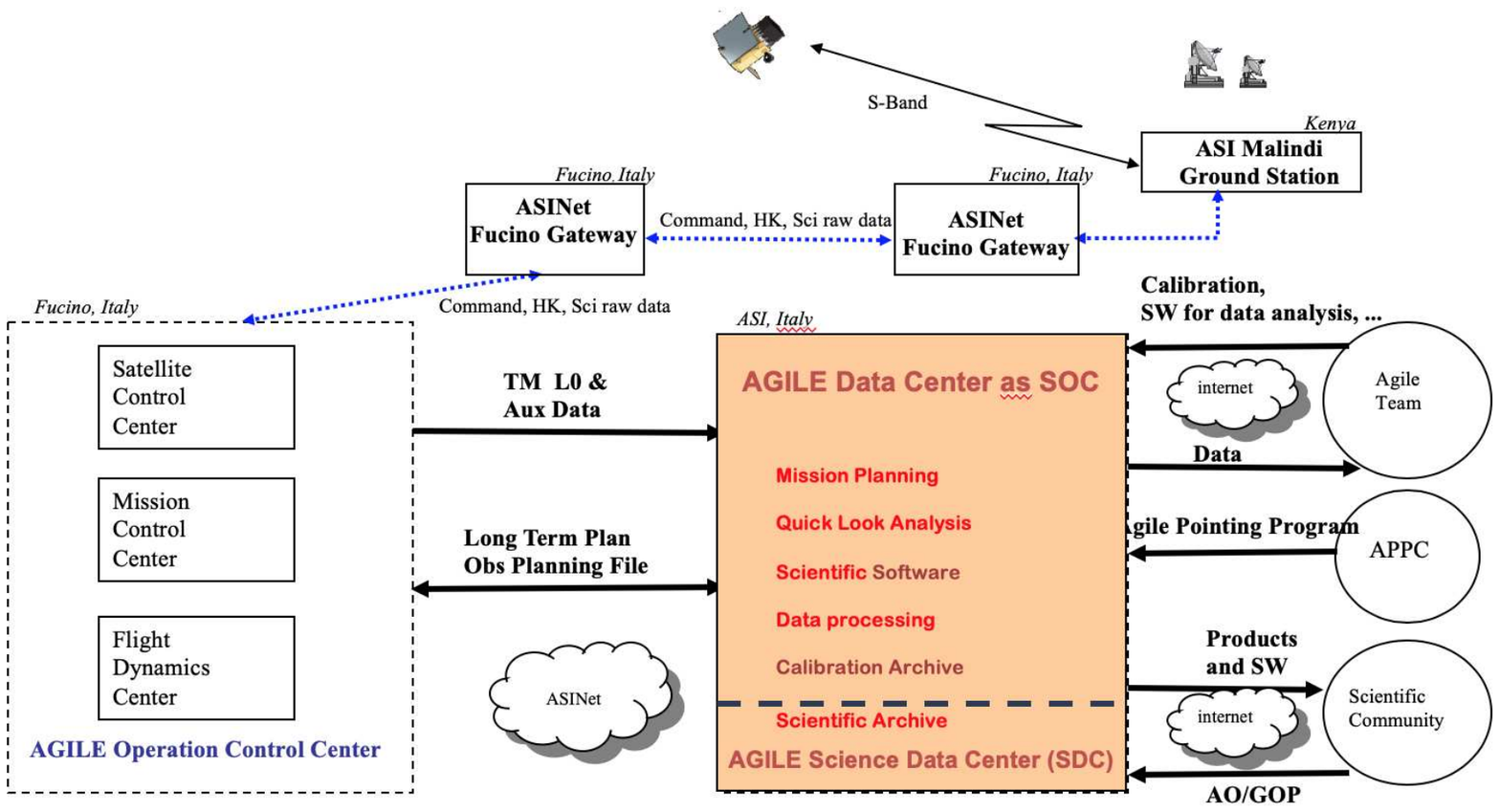}}}
\vskip -4.8 truecm
\caption{The Ground Segment (GS) architecture of the AGILE space mission.
The AGILE Data Center acts both as a Science Operation Center (SOC) and as 
a Science Data Center (SDC).}
\label{fig02}
\end{figure*}

In the context of the Ground Segment (GS) architecture of the mission,
the AGILE Data Center acts both as a Science Operation Center (SOC)
and a Science Data Center (SDC), see scheme in Fig. \ref{fig02}.

The main ADC activities and responsibilities include: 
scientific requirements implementation,
development and implementation of data 
processing systems (pipelines), 
relational databases and data archives. 
Management of the satellite observations include:
real time data
acquisition and monitoring, alert systems, processing (and reprocessing), 
archiving, standardization, analysis
and interpretation of data, software tools development to grant optimal 
diffusion, accessibility
and usability of scientific data.
ADC also managed the Guest Observer (GO) Program of the AGILE mission, 
and it is responsible
for the standardization, inclusion and distribution of public scientific data  
in the large SSDC interactive 
multi-mission archive.
More details are given in the following sections.

\section{The AGILE Data Flow}
\label{FLOW}

\begin{figure*}
\vskip -0.5 truecm
\centerline{\resizebox{120mm}{!}{\includegraphics[angle=-90]{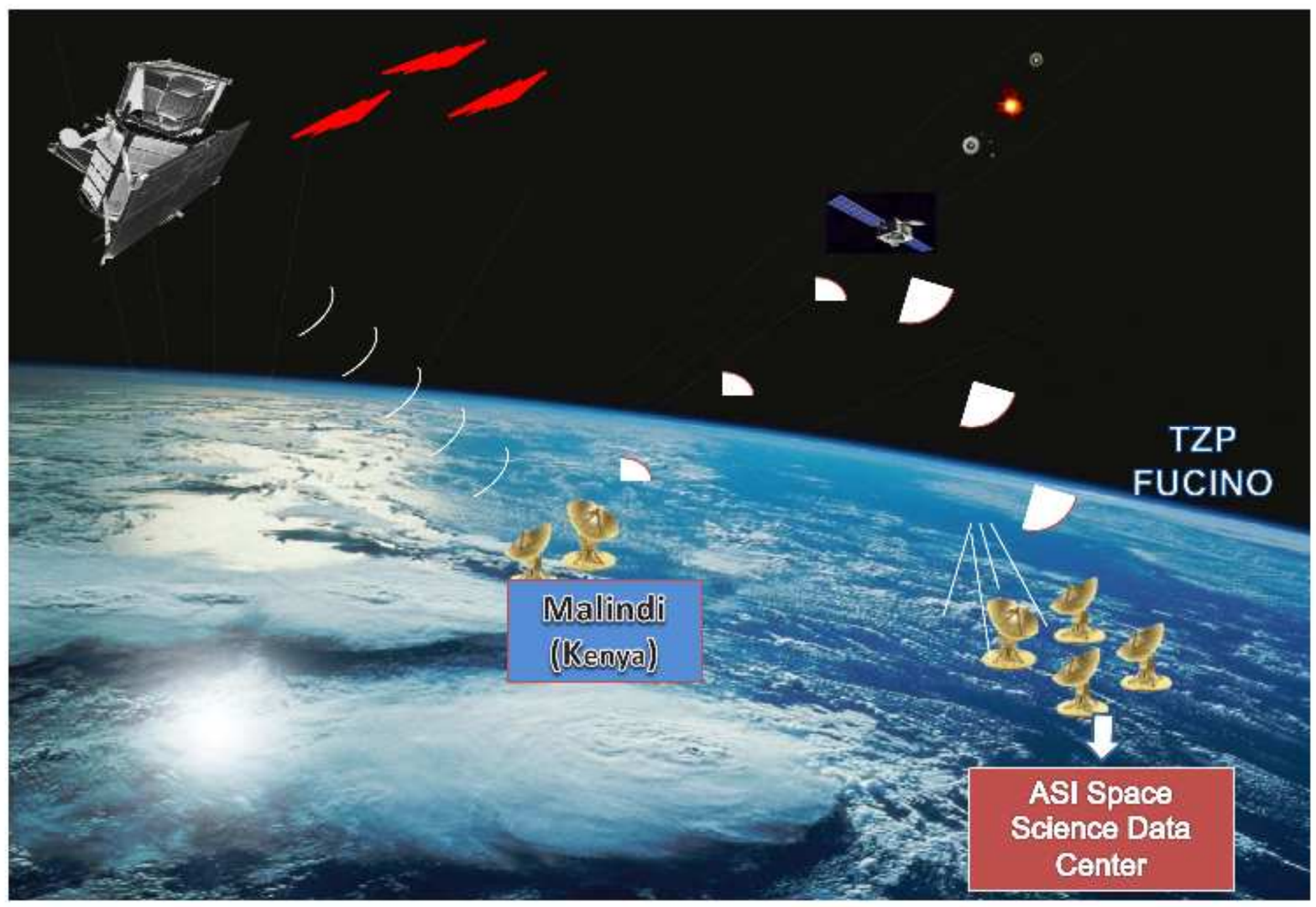}}}
\vskip -0.5 truecm
\caption{Pictorial view of the data flow of the AGILE Ground Segment.}
\label{fig03}
\end{figure*}

AGILE raw telemetry level-0 data (LV0)
are down-linked every {$\sim$} 100 min to the ASI Malindi 
ground station in Kenya and transmitted, through
the fast ASINET network provided by ASI, first to the Telespazio 
Mission Control Center at Fucino, and then to the ADC within 
{$\sim$} 5 min after the end of each contact downlink, see Fig. \ref{fig03}.

The AGILE-GRID ground segment alert system
is distributed among ADC and the AGILE Team Institutes and it
combines the ADC quick-look
with the AGILE Science Alert System
developed by the AGILE Team \cite{Bulgarelli2014}. 
GRID Alerts are sent via email (and sms) both on a
contact-by-contact basis and on a daily timescale.
The already very fast AGILE ground segment alert system 
(with alerts within $\sim$ (2-2.5) hours 
since an astrophysical event)
has been further optimized 
in May 2016,
for the search of electromagnetic counterparts of gravitational waves
and other transients. Currently the system allows the AGILE Team 
to perform a full data reduction 
and the preliminary Quick Look (QL) scientific analysis only 25/30 minutes 
after the telemetry download from the spacecraft.

Refined analysis with human intervention of
most interesting alerts are
performed every day (quick-look daily monitoring).

\section{AGILE observation modes}
\label{obsmodes}

\begin{figure*}
\vskip -4.7 truecm
\centerline{\resizebox{120mm}{!}{\includegraphics{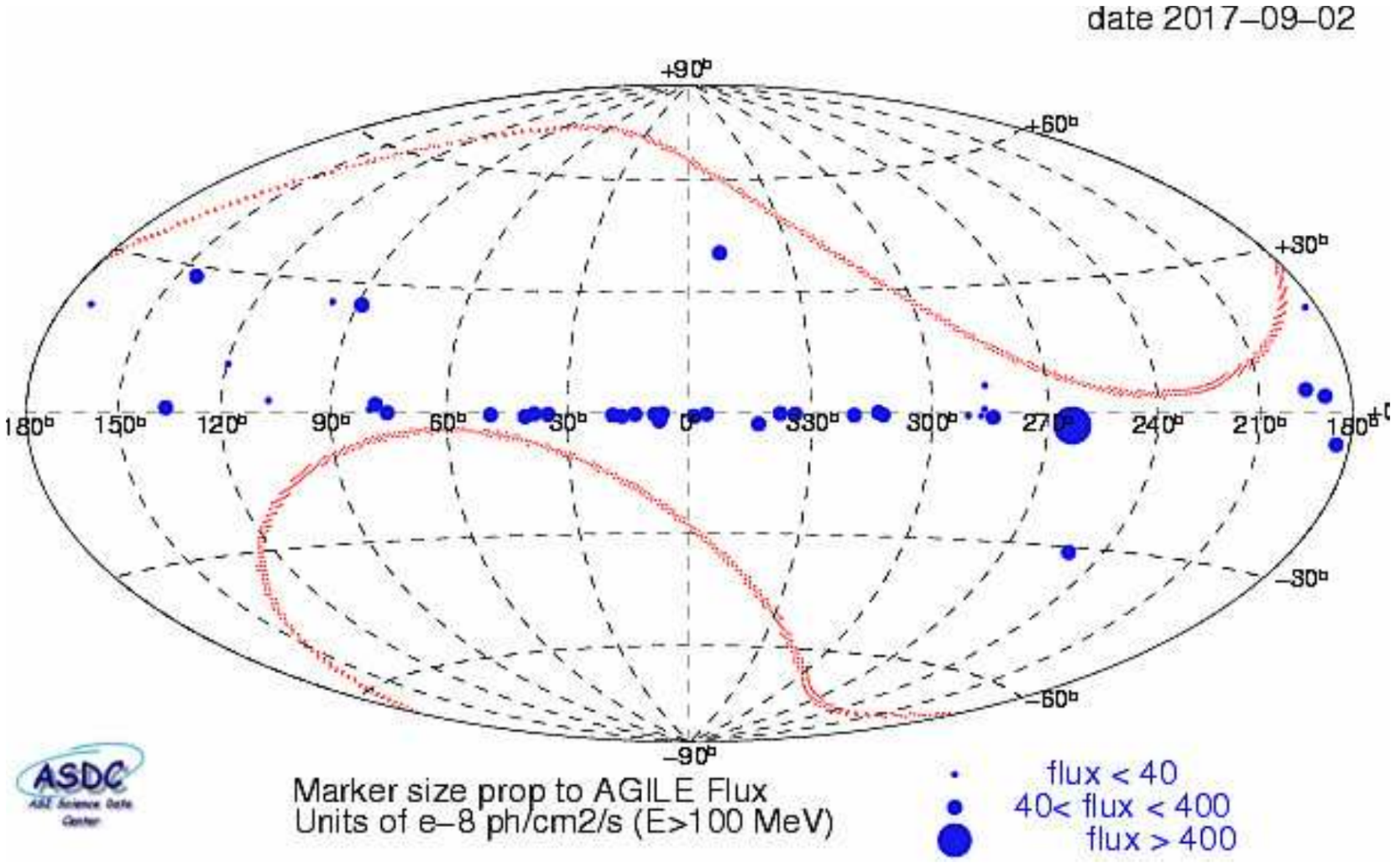}}}
\vskip -4.7 truecm
\caption{The large AGILE-GRID FoV in spinning on a particular day. The sky regions not
accessible for solar panel constraints are delimited in red, and indicate the mean daily sun
and anti-sun positions.}
\label{fig04}
\end{figure*}

\subsection{Pointing mode}
The AGILE pointings are subject to illumination constraints requiring that
the fixed solar panels always be oriented within about 3 degrees from the
Sun direction. During the first $\sim$ 2.5 years AGILE was operated in ``pointing
observing mode'', characterized by long observations called Observation Blocks
(OBs), typically of 2-4 weeks duration, mostly concentrated along the Galactic
plane following a predefined Baseline Pointing Plan. The AGILE Pointing
Plan during the first two years of observations (Cycle-1 and Cycle-2) has been
prepared taking into account several scientific and operational requirements
such as: maximisation of the overall sky exposure by limiting the observation
of the sky regions more affected by Earth occultation; substantial exposure
of the Galactic plane, in particular of the Galactic Center and of the Cygnus
region, in order to achieve good statistic for several $gamma$-ray pulsar candidates,
micro-quasars, supernova remnants, and unidentified $gamma$-ray sources with
simultaneous hard X-ray and $gamma$-ray data.
During the pointing period phase, 
even though the exposure was mainly focused
towards the Galactic plane, the large AGILE FoV has allowed to achieve a
good balance between Galactic and extra-galactic targets, with fast reaction
capability to transient $gamma$-ray sources.
The list and details of the 101 OB of AGILE observations in pointing mode
can be found at the ADC dedicated webpages\footnote{https://agile.ssdc.asi.it/current$\_$pointing.html}. 

\subsection{Spinning mode}

On November 4, 2009, AGILE scientific operations were reconfigured 
following a malfunction of the rotation wheel occurred in mid October, 2009. 
Since then, the satellite is operating
regularly in ``spinning observing mode'', with
the solar panels pointing at the Sun and the instrument axis sweeping the
sky with an angular speed of about 0.8 degree/sec. The instrument and all
the detectors are operating nominally producing data with quality equivalent
to that obtained in pointing mode. AGILE in spinning mode is surveying a
large fraction (about 80\%) of the sky each day. The AGILE Guest Observer
Program has not suffered any interruption.

\begin{figure*}
\vskip -0.5 truecm
\centerline{\resizebox{120mm}{!}{\includegraphics[angle=-90]{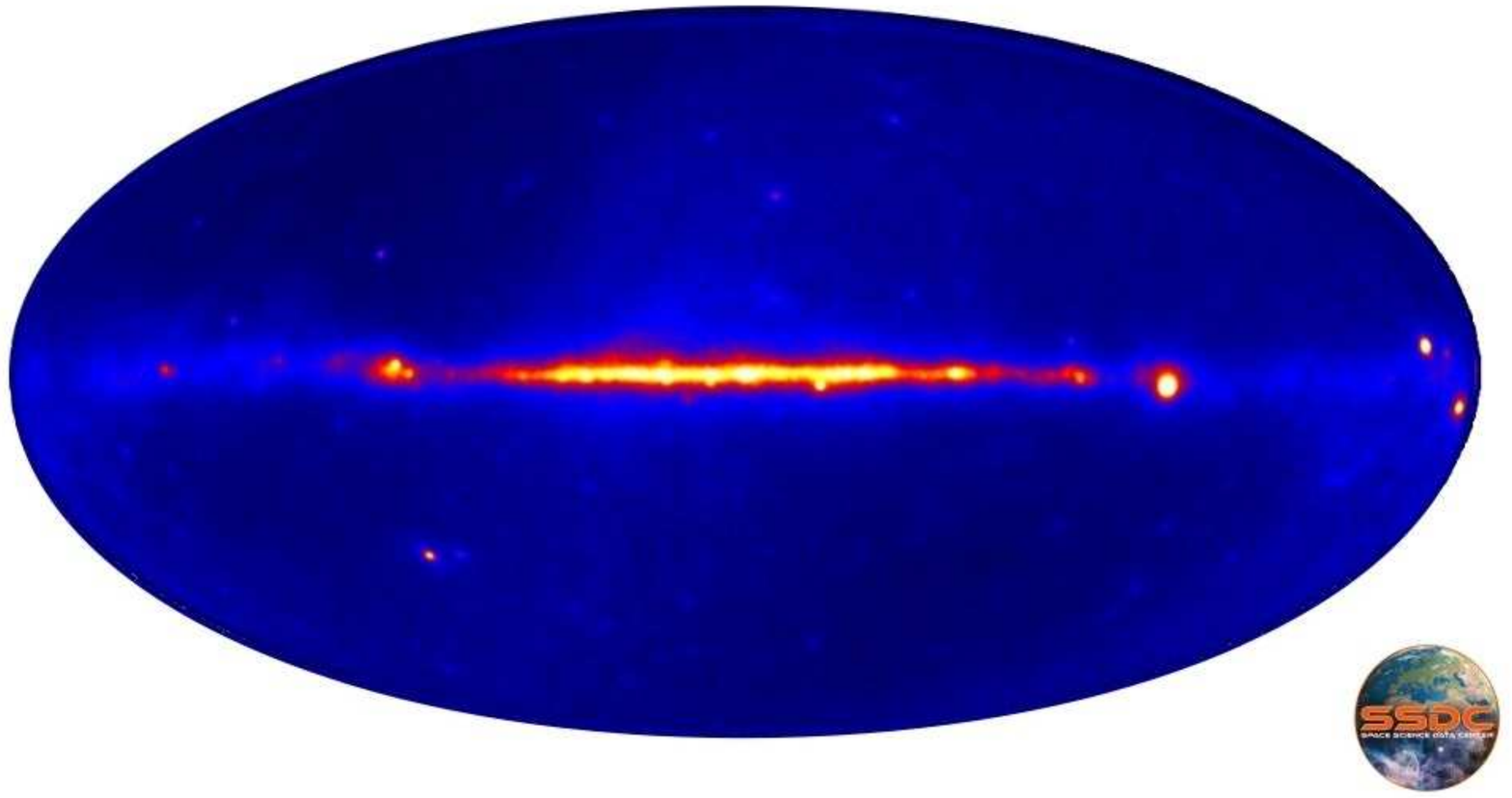}}}
\vskip -1. truecm
\caption{AGILE-GRID total intensity map for energies above 100 MeV (Pointing + Spinning)
from December 1, 2007 up to Sep. 30, 2017.}
\label{fig05}
\end{figure*}

An example of the accessible sky view in spinning on a particular day is
shown in Fig. \ref{fig04} and the total 10-year AGILE intensity map above 100 MeV
from December 1, 2007 up to September 30, 2017 (Pointing + Spinning) is shown in Fig. \ref{fig05}.
AGILE observations in spinning continue to be structured as a series 
of OBs with a ``dummy'' pointing direction, each corresponding to a 
unique identifing number and lasting $\sim$ 15 days.
In spinning mode, the mean pointing direction coordinates do
not have a meaning. Rather ADC provides the mean Sun positions 
that determine the sky regions that are not accessible during each OB
($\sim$ 60 degrees around the Sun and anti-Sun positions).
Note that the allowed istantaneous pointing directions lie on a great circle orthogonal to the Sun direction,
whose orientation changes with time, so that the whole sky is accessible during a six months period.

\section{Data Processing and Data Levels}
\label{PROCESSING}

The AGILE binary LV0 TM received at ADC is automatically archived after a consistency check, 
and transformed in level-1 FITS format (LV1) through the AGILE Pre-Processing System
(TMPPS) \cite{Trifoglio2008}. 

The Level-1 data are then processed using the scientific data reduction 
software tasks developed by the AGILE
instrument teams and integrated into an automatic 
quick-look pipeline system developed at ADC.

\subsection{ADC Quick look: real time data processing} 

As soon as the Level-1 FITS files of the AGILE contact are produced by the TMPPS
and archived, the data processing involves the following steps:
\begin{itemize}

\item[1)] A first step (CORRECTION) on a contact-by-contact basis, aligns times all Telemetry 
to Terrestrial Time (TT) and performs some preliminary calculations and unit conversions. 
The resulting archive of FITS files is called LV1corr.

\item[2)] In a second step (QLSTD) for GRID Telemetry Data 
an ad-hoc implementation of the Kalman Filter technique 
is used for track identification and event direction reconstruction in detector coordinates. 
Subsequently, a quality flag is assigned to each GRID
event: (G), (P), (S), and (L), depending on whether it is recognized as a $\gamma$-ray event, a
charged particle event, a single-track event, or if its nature is uncertain (limbo), respectively. 
An AGILE auxiliary data file (LOG file) is created, 
containing all the information relevant to the computation of
the exposure, live-time, orbital and attitude reconstruction.
Then the AGILE event files are created, excluding background events
flagged as particles, also reconstructing the event direction in sky coordinates.
A QLSTD process, recently optimised in order to reduce latency, starts 
as soon as all the necessary input telemetries (GRID, Star Sensors, GPS, 
Housekeepings) have a common minimum time coverage.

\item[3)] As a third step, the ADC Scientific Quick Look Analysis (QLSCI) is run. 
Counts, Exposure, and Galactic background $\gamma$-ray maps are created
populating the QL Level-3 (LV3) archive.
To reduce the particle background contamination 
only events flagged as confirmed $\gamma$-ray events
are selected, excluding observations during the South Atlantic Anomaly (SAA).
To reduce the $\gamma$-ray Earth Albedo contamination, 
events whose reconstructed directions form angles with the satellite-Earth 
vector smaller than 80 degrees are rejected.
The ADC QLSCI processing results in a set of Daily Reports with selected candidate detections
which are sent via email to the AGILE Team twice a day.
\end{itemize}

The ADC results of the steps 1) and 2) 
are also promptly forwarded to the AGILE Team site at INAF-OAS in Bologna (previously known as IASF-BO),
where the AGILE Team Science Alert System (SAS) pipeline runs, generating alerts on a contact-by-contact
basis with independent flare search algorithms. These alerts are
sent via SMS, e-mail, and through the notification system of 
the dedicated App for smartphones and tablets AGILEScience \cite{Bulgarelli2014}.

\subsection{ADC Standard Analysis and Consolidated Archive} 

At the end of each AGILE Observation Block, the real-time quick-look archive 
used for alert generation, is consolidated by some completeness checks and possible reprocessing. 
Then the ADC Standard Analysis is run on each OB, producing the
official
AGILE-GRID level-2 EVT and LOG archive (LV2STD), 
compliant with the Office for Guest 
Investigator Programs (OGIP) standards recommended for FITS files,
to be used for all the scientific publications, except for GCNs and Atels alerts.
Data to be distributed to the Guest Observers are extracted from LV2STD event files,
including $\gamma$-ray events in a region of 15 degrees radius around the requested source positions.


\section{The ``AGILE Services'' and the AGILE Storage at ADC}
\label{agileservices}

\begin{figure*}
\vskip -1 truecm
\centerline{\resizebox{120mm}{!}{\includegraphics[angle=-90]{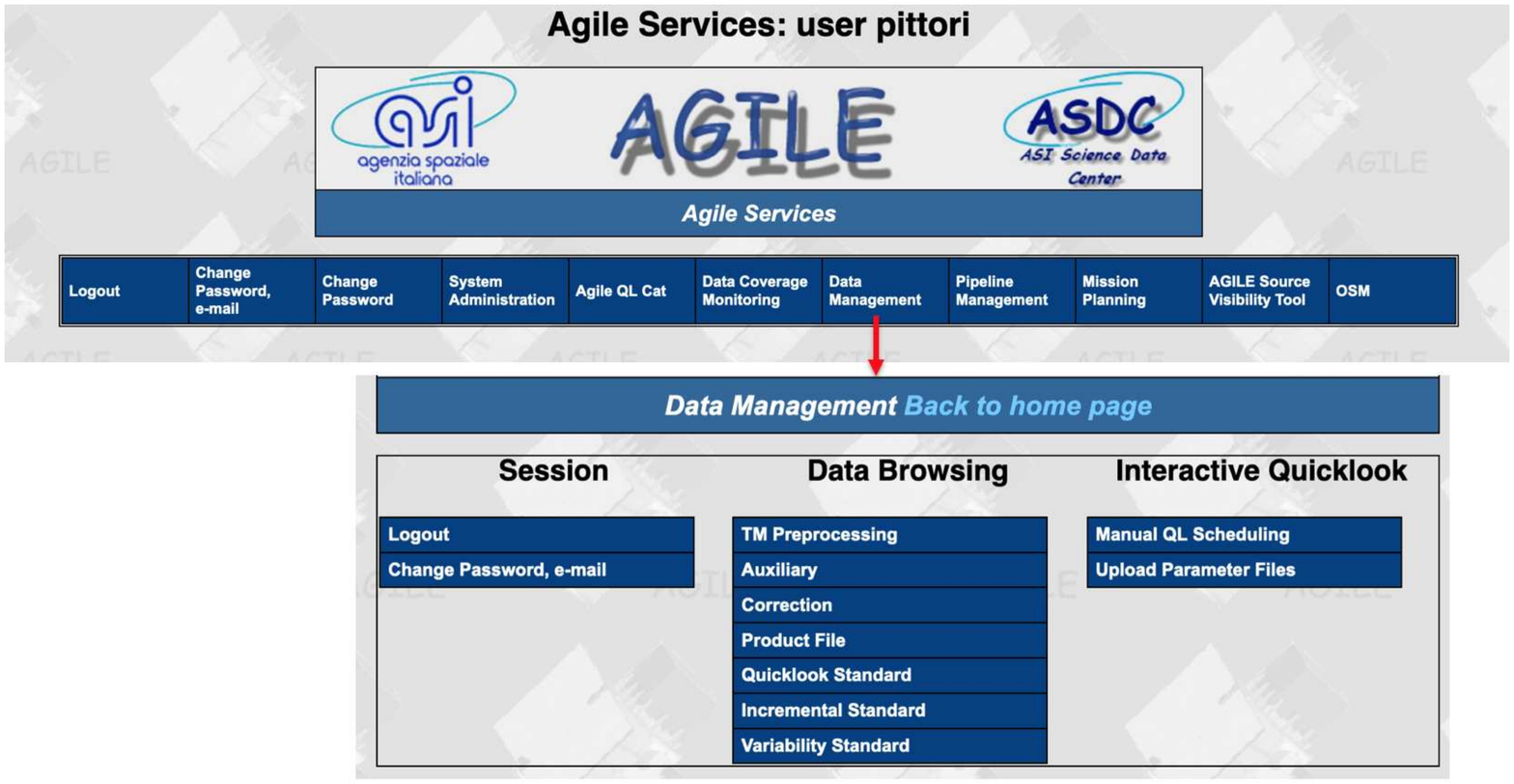}}}
\vskip -1.2 truecm
\caption{A screenshot from the ``AGILE Services'':
a web interface to the Mission Flexible Database relational DB.}
\label{fig06}
\end{figure*}

Management and control of processing pipelines
data and the corresponding outputs is entrusted to a management 
system developed by the IT support of SSDC, 
including the Common Pipeline Control Subsystem.
A Flexible Database Web Interface (password protected) 
to the AGILE Mission relational database (MySQL DBMS), 
called ``AGILE Services'', allows to
represent hierarchical relationships between the data,
to get information about the automatic pipeline processes,
to navigate between the connected data, to read, insert,
modify, and retrive archived data.
Different kind of accredited users (administrators, internal ADC operators, AGILE Team scientist, Guest Observer)
have access to different functionalities: Data Mangement, Pipeline Management, Mission Planning, etc.,
see Fig. \ref{fig06}.
A Proposal Management and Data Distribution system has been specifically developed for the AGILE Guest Observers,
including an AGILE Target Visibility Computation Tool.

The AGILE data storage at ADC increase at a rate of about 1 TB/year. The AGILE consolidated archive, including
reprocessing and QL data contains about 10 TB at the time of writing.

\section{AGILE Data Policy and Data Distribution}
\label{datapolicy}

For the first seven years all AGILE data (i.e. both from the AGILE Team Projects and from the Guest
Observer Program) have been subject to the proprietary rules normally applied to
observatory space data: one year proprietary period after which they
have been available via the public AGILE Data Archive at the SSDC. The one-year
proprietary period started from the date when the Guest Observer or the AGILE Team
received the data in a format that is suitable for analysis and publication.

During the extended lifetime of the AGILE mission, a change in AGILE gamma-ray scientific data policy, 
proposed by the Mission Board, has been approved by ASI to strengthen the engagement 
of the Scientific Community by eliminating the one year proprietary period requirement.
Starting from October 2015 all AGILE-GRID data are published as soon as they are 
processed and validated.
The new public AGILE archive now contains all data from December 1, 2007 up to May 31, 2019, i.e. from 
OB 4900, start of Cycle-1 up to OB 32000 of the on-going Cycle-12.
Public AGILE data are available from the ASI Space Science Data Center
Multimission Interactive Archive (MMIA) webpages.

\subsection{Public Scientific Software}
For refined GRID  scientific analysis a new AGILE public scientific 
software package (AGILE\_SW\_6.0) adapted from the 
AGILE Science Tools developed by the AGILE Team 
(TAGNAME = BUILD25) is available from the ADC webpages. 
It includes new scientific software tasks and calibrations, 
an updated model for the Galactic diffuse gamma-ray emission, 
a refined procedure for point-like source detection, 
and the search for extended gamma-ray sources.

\section{AGILE Catalogs and ADC Interactive Webpages}
\label{catalogs}

\begin{figure*}
\vskip -0.5 truecm
\centerline{\resizebox{110mm}{!}{\includegraphics[angle=-90]{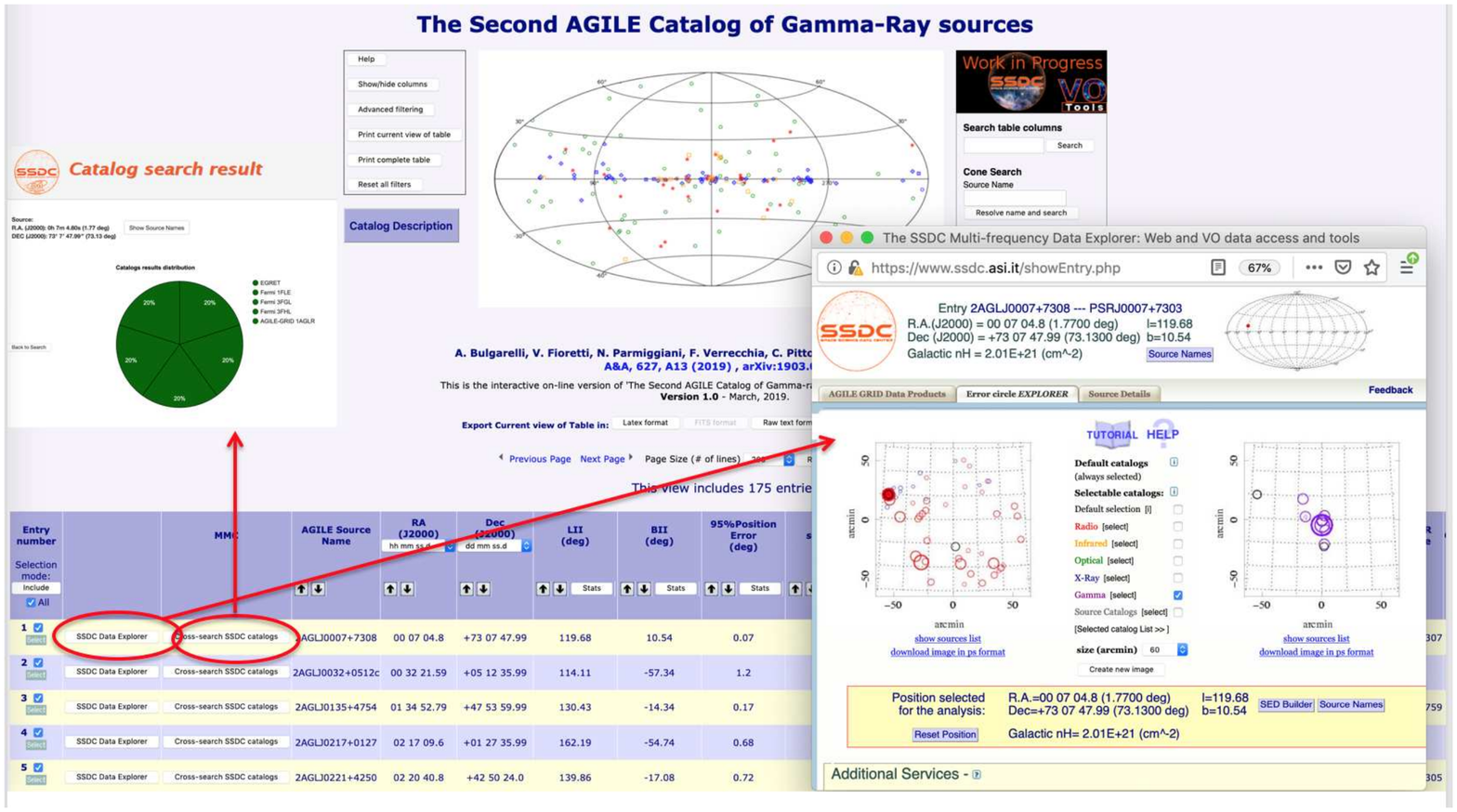}}}
\vskip -1. truecm
\caption{Online version at SSDC of the Second AGILE Catalog of Gamma-Ray Sources \cite{Bulgarelli19}.
The interactive web table allows to access additional data, including AGILE skymaps, the broad band spectral energy distributions through the SSDC SED-Builder tool and the finding charts through the SSDC Data Explorer tool.
}
\label{fig07}
\end{figure*}

Nine AGILE catalogs have been published for the AGILE Mission at the time of writing 
this paper:
four AGILE-GRID $\gamma$-ray source catalogs \cite{Pittori09,Verrecchia13,Rappoldi16,Bulgarelli19},
three AGILE-MCAL Terrestrial Gamma-Ray Flashes (TGF) catalogs \cite{Marisaldi14,Marisaldi15,Marisaldi19},
one AGILE-MCAL GRB catalog \cite{Galli13}, and one SuperAGILE source catalog \cite{Feroci10}.
For each published AGILE catalog an online ADC version is available
in the form of interactive SSDC webtables, e.g. Fig. \ref{fig07}.
The interactive ADC catalog tables, often providing supplementary material to the published papers,
have many functionalities and also provide access
to the multimission SSDC scientific tools (cross-search, sky-region data explorer, 
spectral energy distribution tool, etc.). All the catalogs webpages are also being 
standardised for Virtual Observatory functionalities.

\section{AGILE Legacy Archive and the AGILE-LV3 Web Tool}
\label{LV3}

The $\gamma$-ray scientific analysis over long time scales may require long
processing times. For example, to produce deep exposure maps over a time interval of about
seven months, centered on the Crab Nebula position, can take at least 2 hours on 3 GHz Xeon
CPUs. To speed up the AGILE-GRID scientific analysis, a complete level-3 (LV3) archive of pre-compiled 
exposure, counts and diffuse background maps over 1-day integration time, 
with standard parameters was created at ADC.
The centers of the maps, their size and other relevant parameters are fixed, and have been
chosen in such a way as to cover the whole sky.
This AGILE-LV3 legacy archive can be used as basis for scientific Maximum Likelihood 
analysis on time scales that may vary from weeks to months, or even over the entire duration of the mission.

\begin{figure*}
\vskip -1 truecm
\centerline{\resizebox{110mm}{!}{\includegraphics[angle=-90]{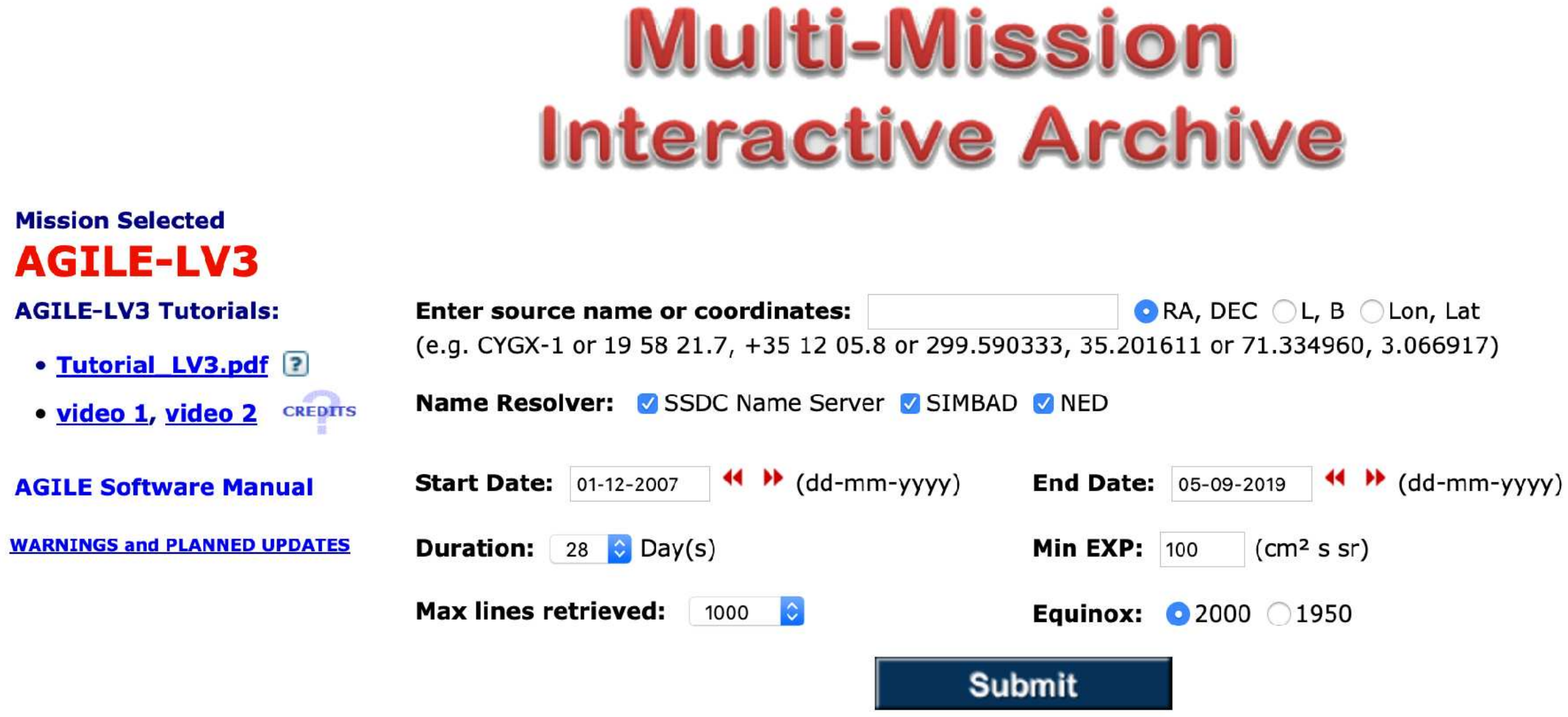}}}
\vskip -1.5 truecm
\caption{The SSDC MMIA query page for the AGILE-LV3 scientific analysis web tool.}
\label{fig08}
\end{figure*}

For an easy on-line AGILE official data analysis, the interested user may query the entire public LV3
archive through the AGILE-LV3 data analysis web tool, Fig. \ref{fig08}.
In the query page the user can enter the source name or sky coordinates 
of the object he/she wants to analyze, the period of interest and the duration 
of the LV3 maps (e.g. 1, 7 or 28 days timebins) to be used in the analysis.
The output from the query automatically selects all AGILE available observations of the source.
The user can execute the standard AGILE Maximum Likelihood (ML) analysis by clicking on 
the Interactive Analysis buttons, or directly generate the $\gamma$-ray light curve over the selected
period, with waiting times ranging from a few seconds to a few minutes (depending on the number of selected timebins).
The AGILE-LV3 tool is meant to be easily comprehensible, does not require 
any install-on-premises software or calibrations, and it has been also tested with high-school students.

\begin{acknowledgements}
This paper in written on behalf of the 
scientific team of the AGILE Data Center at SSDC, which is currently composed by:
C. Pittori (coordinator), F. Lucarelli, and F. Verrecchia (deputy coordinator), with the main IT support of 
G. Fanari (Telespazio).

Many people have contributed to the ADC in the past 10 years,
including scientific staff and IT support staff, with variable FTEs
depending on the operational phase of the Mission.
We acknowledge the contribution of (in alphabetical order):
A. Antonelli, S. Cutini, S. Colafrancesco, D. Gasparrini, P. Giommi, M. E. Pennisi, B. Preger, P. Santolamazza
(scientific support); F. Acerra, P. D'Angeli, A. Guerra, D. Navarra, W. Oliva, R. Primavera, S. Stellato,
F. Tamburelli (IT support).

The ADC operate in close relationship with the Telespazio
Mission Operation Center at Fucino, and in particular with the 
spacecraft operations manager (SOM) P. Tempesta.

We also thank the AGILE Mission Board, composed of: the
PI of the AGILE Mission M. Tavani, the Co-I G. Barbiellini,
the current ASI Mission Director F. D'Amico, 
as well as the former Mission Directors: Luca Salotti, up to September 20, 2010, 
and Giovanni Valentini up to January 22, 2015, and the ASI representative E. Tommasi di Vignano.

We would like to acknowledge the financial support of ASI
under contract to INAF: ASI 2014-049-R.0 dedicated to SSDC.

The paper is dedicated to the memory of the late Francesca Tamburelli.

\end{acknowledgements}

\section*{Compliance with Ethical Standards}
The author declares that there is no conflict of interest.



\end{document}